\documentclass[preprint,showpacs,aps,prc,groupedaddress,floatfix]{revtex4}
\usepackage{graphicx}

\usepackage{amsmath,amssymb} \usepackage{bm}

\newcommand{\hlf}{\mbox{$\frac{1}{2}$}}

\newcommand{\beq}{\begin{equation}} 
\newcommand{\eeq}{\end{equation}}

\def\nuc#1#2{\relax\ifmmode{}^{#1}{\protect\text{#2}}\else${}^{#1}$#2\fi}
\begin{document} \graphicspath{{figures/}}

\title{The dynamic polarization potential and dynamical non-locality in nuclear
potentials: Deuteron-nucleus potential}

\author{R. S. Mackintosh} \email{raymond.mackintosh@open.ac.uk}
\affiliation{School of Physical Sciences, The Open University, Milton
Keynes, MK7 6AA, UK} \author{N. Keeley} \email{nicholas.keeley@ncbj.gov.pl}
\affiliation{National Centre for Nuclear Research, ul Andrzeja So\l tana 7, 05-400
Otwock, Poland} \date{\today}

\begin{abstract} 
The consequences for direct reactions of the dynamical non-locality
generated by the excitation  of the target and projectile are  much less
studied than the effects of non-locality arising from  exchange
processes. Here we are concerned  with the dynamical non-locality due to
projectile excitation in deuteron induced reactions. The consequences of
this non-locality can be studied by the comparison of deuteron induced
direct reactions calculated with alternative representations of the
elastic channel wave functions: (i) the elastic channel wave functions
from coupled channel (CC) calculations involving specific reaction
processes, and, (ii) elastic channel wave functions calculated from
local potentials that exactly reproduce the elastic scattering
$S$-matrix from the same CC calculations. In this work we produce the
local equivalent deuteron potentials required for the study  of direct
reactions involving deuterons. These will enable the study of the
effects of dynamical non-locality following a method  previously
employed in an investigation of the effects of non-locality due to
target excitation. In this work we consider only excitations due to
deuteron breakup, and some new properties of the breakup dynamical
polarization potential  (DPP) emerge that reveal dynamical non-locality
directly. In addition, we evaluate the TELP inversion method and find
that it fails to reproduce some features  of the DPP due to breakup.

\end{abstract}
% pacs corrected 15th Jan after sending to NK

\maketitle

\section{INTRODUCTION}\label{intro} Dynamical non-locality is present in
nucleus-nucleus interactions as a result of coupling between the elastic channel
and inelastic and reaction channels. It is a property of the non-local dynamical
polarization potential, DPP, that is generated by the coupling~\cite{satchler}.
It is much less studied than exchange non-locality and its effects are not
generally accounted for in the DWBA analysis of direct reactions. The inclusion
of dynamical non-locality in DWBA calculations is not
amenable to relatively simple corrections that are applied  to exchange
non-locality. This and other aspects of dynamical non-locality are discussed in
Ref.~\cite{KM14a} which introduces a method for including it in direct reaction
calculations without requiring the solution of integro-differential equations.
The method was used to study the dynamical non-locality of the nucleon optical
potential  that arises from coupling to collective states of the target nucleus.
This was accomplished by comparing angular distributions for nucleon transfer
reactions involving two alternative representations of the nucleon optical model
potential, OMP. The alternative nucleon potentials were: (i) a potential having
dynamically induced non-locality, and, (ii) a local potential having an
identical partial wave S-matrix $S_l$ (and hence identical elastic scattering
observables) for all values of the partial wave orbital angular momentum $l$. 
The local potential was derived from the $S$-matrix of the
non-local potential by $S_l\rightarrow V(r)$ inversion.  The details are in
Ref.~\cite{KM14a}. 

The same general approach can be applied to the study of the non-locality
arising from the excitation of a projectile rather than the target. The case of
deuteron scattering is of particular interest since, of all nuclei, its cluster structure
is best understood. Moreover, deuteron induced reactions are a very important
 source of spectroscopic information. In a subsequent paper we will study the effect
 of dynamically induced non-locality in various deuteron induced reactions:
 ($d,t$), ($d,p$) and ($d,\nuc{6}{Li}$).  The dynamical non-locality is generated by
coupling to breakup states as the deuteron interacts with a target nucleus. The
consequences of this non-locality can then be evaluated by comparing angular
distributions for ($d,p$),  and other deuteron-induced reactions, calculated  (i) with
dynamically non-local deuteron OMPs, and (ii) with their local equivalents. The local
deuteron OMPs have the same S-matrix and hence elastic scattering observables as
the dynamically non-local OMPs. This procedure will not incorporate the complete 
effects of deuteron breakup in ($d,p$) reactions, see Ref.~\cite{JandS,JandT}, just 
the effect of the dynamical
non-locality that would be relevant to any direct reaction involving deuterons,
($d,t$) and ($d,\nuc{6}{Li}$) for example.  Rather complete descriptions of
breakup and exchange processes  specifically for ($d,p$) and ($p,d$) 
reactions exist~\cite{nunes}.  However, the emphasis in this work is different, being a general study of 
dynamical non-locality arising from projectile excitations, based on deuterons as a projectile. The
effects of this dynamical non-locality on a range of transfer reactions can then be studied
applying the procedure of Ref.~\cite{KM14a}.

Studying the effect of the dynamical non-locality induced by deuteron breakup is motivated in part 
by the fact that non-locality due to other inelastic processes gives rise to effects on the projectile 
wave function that are very different from the standard Perey effect. These effects were studied in 
Ref.~\cite{mic88} and Ref.~\cite{cm90}. The first of these found an `anti-Perey` effect in a case of 
rotational coupling, and the more systematic study in the latter defined a `generalised Perey factor',
GPF. In a case where the standard Perey factor would be a uniform 0.85 within the nucleus, this was 
found to be just the case for Perey-Buck non locality. However, for the case of inelastic scattering 
involving the excitation of vibrational states of the target, the GPF exhibited a complex pattern of 
regions where it was greater and less than unity. The GPF being $>1$  corresponds to an `anti-Perey' 
effect. Effects of this kind are likely to modify the wave function of any projectile involved in 
a direct reaction. This could therefore make a significant difference
for any direct reaction involving that projectile.  For this reason the generalised 
`Perey'  effects due to projectile breakup should be tested directly by means 
of ($d,t$), ($d,\nuc{4}{He}$),  ($d,p$), ($d,n$)and ($d,d^\prime$), etc. reactions.

The non-local DPP that is generated by channel coupling is also $l$-dependent,
having a different function of two radial coordinates for each $l$.  However,
since $l$-dependence and non-locality are hard to disentangle, we generally
refer just to `non-locality'. The non-local potentials generated by channel coupling 
are hard to interpret~\cite{rawit87} and the calculation of scattering observable from them
requires the solution of integro-differential equations. Simple recipes, like
the introduction of a Perey factor, are not available, motivating
the procedure presented in
Ref.~\cite{KM14a}. In that work, the projectile wave function is not calculated
from a non-local potential but calculated \emph{in situ}, as it is
generated by coupling, using the coupled channel  code FRESCO~\cite{Tho88}.
We follow the same general procedure for non-locality due to projectile excitation.

In this first paper, we present the breakup calculations from which we determine the
local equivalent potentials and the corresponding DPPs.  These  
local equivalent potentials give exactly the 
same  elastic scattering S-matrix,  $S_l$, and hence exactly the same elastic 
scattering angular distributions, as those resulting  when  deuteron breakup  channels
are coupled to the elastic channel. These  potentials are determined by exact
inversion of the elastic scattering $S$-matrix, $S_l$,   from  specific  breakup calculations. 
We demonstrate that the breakup calculations presented here already provide evidence 
for dynamical non-locality in deuteron-nucleus scattering.

In a subsequent paper~\cite{paper2} we determine the angular distributions   from
a range of direct reactions all calculated with dynamical  non-locality in the
deuteron channel. To evaluate the effect  of dynamical non-locality, 
these angular distributions  will be compared with angular distributions calculated using  the local deuteron 
potentials that are determined in the present paper. In this work it has been possible to exploit
the  calculations presented here to go beyond supporting the evaluation of dynamical
non-locality in the subsequent paper. Section II details the deuteron breakup calculations 
and the inversion leading to local DPPs; Section III presents the DPPs for various classes of breakup
and discusses their various generic properties; Section IV presents further implications including 
direct evidence of non-local effects; Section V evaluates and also exploits the trivially equivalent local 
potential, TELP, inversion method; Section VI is a general discussion of our findings and the Appendix formally supports an argument in Section IV concerning implications of non-locality.

Throughout this work we use $l$ for partial wave orbital angular momentum and
$L$ for the relative orbital angular momentum of the nucleons
in a deuteron. Spin is not included in the present calculations.

\section{BREAKUP CONTRIBUTIONS TO DEUTERON-NUCLEUS  INTERACTIONS}\label{method} The
interaction of a deuteron with a nucleus is strongly modified by the breakup of
the deuteron. The coupled discretized continuum channel, CDCC,  method for calculating breakup is now well  
established~\cite{rawit74,farrell76,yahiro82,austern87,Sakuragi-PTPS-1986}. 
Although CDCC is not rigorous, it is reasonable for the purpose of the present study. 

To study the importance of this dynamical non-locality, we require the local deuteron
potential that is S-matrix equivalent to the non-local potential generated by coupling 
to the breakup  states of the deuteron. Such breakup states implicitly
contribute~\cite{MK82}  to the empirical local deuteron OMP; for a recent study see 
Ref.~\cite{pang}. The local deuteron potential, that we call `S-matrix 
equivalent',  is determined by $S_l\rightarrow V(r)$ inversion 
to have exactly the same S-matrix as the elastic channel S-matrix from the 
CDCC breakup calculations. A local and $l$-independent representation of the 
dynamic polarisation potential (DPP) is found by subtracting  the elastic 
channel potential of the breakup calculation (the `bare' potential) from 
the potential found by inversion.

\subsection{The deuteron breakup calculations}\label{BUcalc} 
In the subsequent paper, Ref.~\cite{paper2}, we  study direct transfer reactions involving 
30 MeV deuterons incident upon
\nuc{16}{O} and we present here the relevant CDCC  breakup calculations. These calculations
were carried out with the coupled channel  code  {\sc Fresco}~\cite{Tho88} which is
also used in the  transfer calculations with dynamically non-local interactions~\cite{paper2}. 
The deuteron and
its constituent nucleons were all considered to be spinless for the sake of simplicity. The
neutron-proton binding potential was of Gaussian form and was taken from Ref.\ \cite{rawit74}.  
Coupled discretized continuum channels (CDCC) calculations, similar to those described in
Ref.\ \cite{Raw80}, were performed, the $n+^{16}$O and $p+^{16}$O optical potentials
required as input to the Watanabe-type folding potentials being taken from the global
parameterization of Ref.\ \cite{Kon03}. The $n+p$ continuum was divided up into bins in
momentum ($k$) space of width $\Delta k = 0.1$ fm$^{-1}$ up to a maximum value $k_{\mathrm{max}}   
= 0.7$ fm$^{-1}$. Five sets of CDCC calculations were performed, labeled L0, L2, L02, L0andL2 
and L024, respectively. These denote calculations with $n+p$ continua of relative angular
momentum $L=0$ only, $L=2$ only, $L=0$ and $L=2$ with full continuum-continuum coupling,
$L=0$ and $L=2$ with separate couplings between bins with $L=0$ and bins with $L=2$ only
(i.e.\ no couplings between bins with $L=0$ and $L=2$), and $L=0$, $L=2$ and $L=4$ with full
continuum-continuum coupling, respectively. Comparison of the inverted potentials from certain 
combinations of these cases will reveal direct evidence of effective non-locality in deuteron-nucleus 
interactions. We also present the characteristics of DPPs calculated in similar calculations which
show that the general properties are generic.

\subsection{The S-matrix inversion}\label{inversion}
The S-matrix inversion is carried out using the iterative-perturbative, IP,
algorithm~\cite{MK82,ip2,kukmac,arxiv,spedia} applied to the elastic channel
$S_l$ from the deuteron breakup calculation. The IP method provides a local
potential yielding S-matrix elements that are effectively indistinguishable from
those of the non-local potential, and hence leading to indistinguishable
scattering observables. In some cases, the potentials found by inversion have a
degree of undulatory behavior in the surface which becomes more pronounced as
the S-matrix $S_l$ from the inverted potential more closely approaches $S_l$ for the 
non-local potential. A strongly  undulatory character is characteristic of
potentials determined by fitting the S-matrix $S_l$ calculated from potentials
that are known to be $l$-dependent. In fact, potentials found by inverting $S_l$
resulting from the many coupled channels calculations that have been studied \emph{always}
have some degree of well-established undulatory character; DPPs are never smooth
and certainly never proportional to the bare potential. All the inverted
potentials discussed herein have that property.  IP inversion can handle
spin-\hlf\ and spin-1 projectiles, but for the present purposes and to simplify
the deuteron breakup calculations we ignore nucleon spin throughout. An
evaluation of the effects of deuteron non-locality upon the $J$-dependence and
analysing powers of transfer reactions  must therefore await a future extension
of this work. The sensitivity of the scattering to the DPP at various 
values of the deuteron-nucleus separation can be established by notch tests.

An alternative method of inversion determines the trivially equivalent local potential, TELP, Ref.~\cite{telp}.
A form of this potential with appropriate partial wave weighting~\cite{weight} is incorporated 
in the FRESCO code~\cite{Tho88} and  it is this weighted TELP (referred to here as just TELP),  that
is commonly  used to derive DPPs. It has been evaluated against exact IP inversion and has 
been shown to be  deficient~\cite{pang1} in specific situations, and further evidence
is presented below. IP inversion 
leads to a potential that will precisely reproduce the elastic scattering angular distribution from 
the CDCC calculation. Although TELP inversion turns out to be inadequate, in general, as an inversion 
method,  we nevertheless find it useful. With IP inversion, the iterative process
can start from any `starting reference potential', SRP. In studies such as the present, the 
SRP is often taken to be the bare potential of the coupled channel calculation. The DPP can then be calculated
immediately by subtraction. However, when the inverted potential is expected to have some 
undulatory features, as in the present case, it is helpful to start the iterative process from a
potential that can be presumed to be a closer approximation to the target potential.
We therefore carried out inversions using the TELP as the SRP. In most cases, the final converged
potential was very close to the potential found using the bare potential as SRP. In some cases,
the potential was smoother in the surface, eliminating some spurious undulations
in that region. However, the overall undulatory features, extending for deuterons to the nuclear
centre, were essentially identical with each SRP.  We conclude that TELP potentials
are useful as the SRP for IP S-matrix inversion. 

As a by-product of the use of TELP as SRP, we were able extract the DPPs as calculated from
the TELP, and thus evaluate the use of the TELP for determining DPPs. This evaluation of TELP
inversion for the present scattering case is presented in Section~\ref{reTELP}.

\section{DEUTERON BREAKUP: DYNAMICAL POLARIZATION POTENTIALS }\label{dpdt} 
\begin{figure}[htb] \centering
\includegraphics[width=0.7\textwidth,angle=-90]{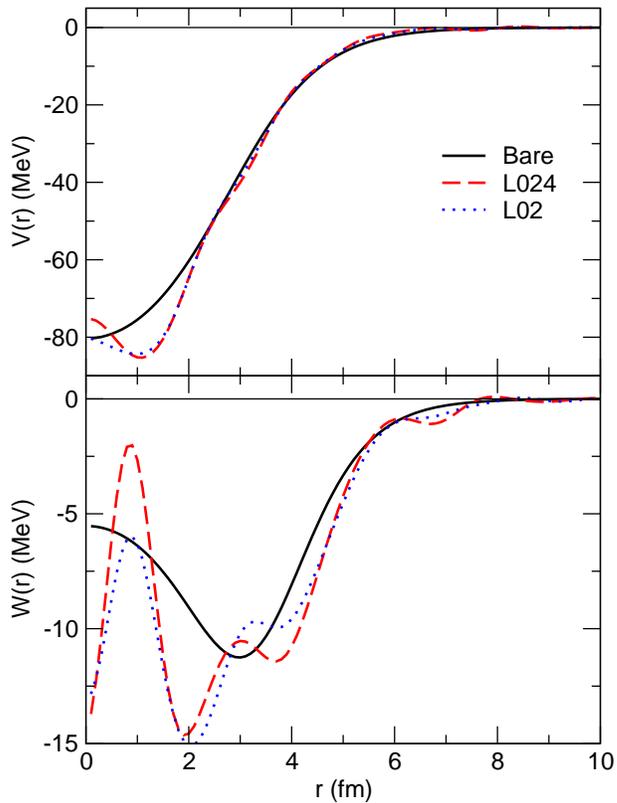}
\caption{\label{fig1}  For 30.0 MeV deuterons on \nuc{16}{O},  showing the contribution
of specific breakup couplings to the real (top panel) and imaginary (lower panel)
elastic scattering interaction. The solid lines are for the bare interaction, the dotted lines
are for the potential found by inversion for the L02 coupling case (defined in the text) and 
the dashed lines show the potential found by inversion for the L024 case.  } \end{figure}

\begin{figure}[htb] \centering
\includegraphics[width=0.5\textwidth,angle=0]{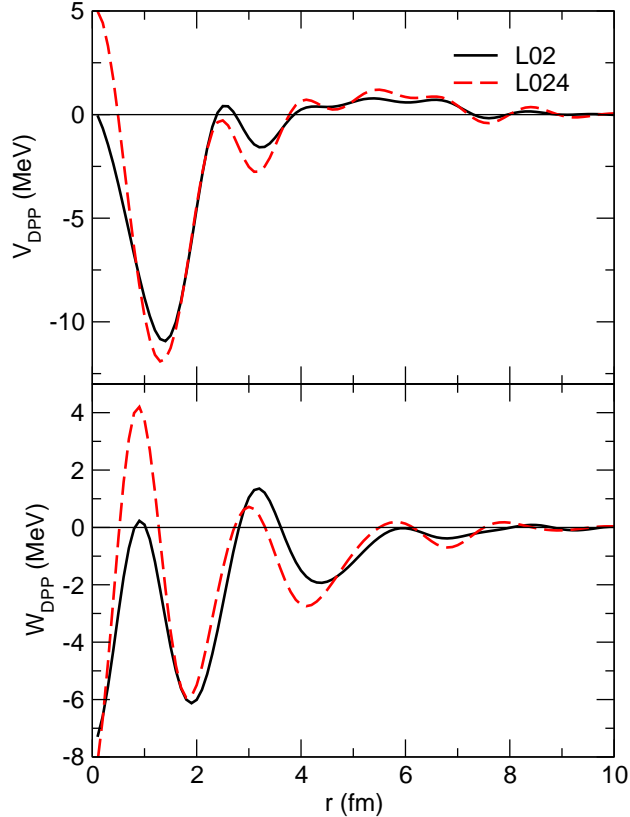}
\caption{\label{fig2}  For 30.0 MeV deuterons on \nuc{16}{O},  showing the real (upper panel)
and imaginary (lower panel) DPPs
for specific breakup couplings. The solid lines give the DPP for the L02 case,  the dashed lines are
for the L024 case.  } \end{figure}

Inversion of the elastic scattering S-matrix $S_l$ that is generated by breakup coupling,
for cases L02 and L024, yields the potentials presented in  Fig.~\ref{fig1} where they are 
compared with the bare potential.   These potentials will form the basis of the determination of 
non-local effects in Ref.~\cite{paper2}. The DPP in each case is found by subtracting the bare 
potential, and the effect of the coupling is more apparent when these DPPs are plotted.
The L02 and L024  DPPs are shown in Fig.~\ref{fig2}. The imaginary DPPs
have emissive regions, more significantly in the L024 case. These properties can be considered 
well established and notch tests reveal that a peak in the sensitivity 
occurs for a notch at around 2 fm, with considerable sensitivity down to 1 fm and less. The 
significant oscillatory features apparent in Fig.~\ref{fig1} and Fig.~\ref{fig2} are therefore 
well within the radial range to which the elastic scattering angular distribution is sensitive. 
The emissive features in the DPP do not generally lead to emissivity in the inverted potential.
An exception can be seen at around 8 fm in Fig.~\ref{fig1}; this cannot lead to $|S_l|$ breaking 
the unitarity limit and $S_l$ for this potential very closely fits the (unitarity respecting) $S_l$\
from the CDCC code.
Establishing such undulatary features is not possible with approximate inversion procedures 
such as the weighted TELP, as we shall show.

\begin{figure}[htb] \centering
\includegraphics[width=0.5\textwidth,angle=0]{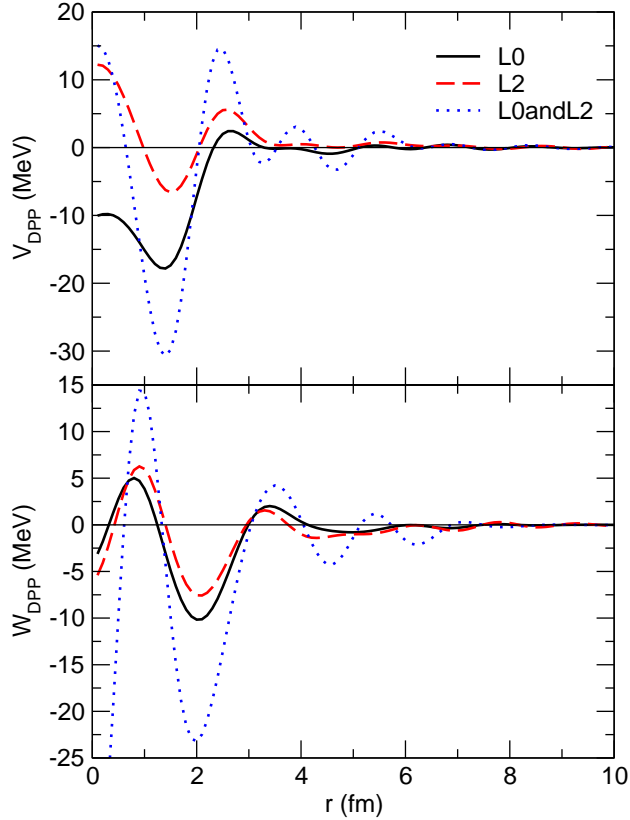}
\caption{\label{fig3}  For 30.0 MeV deuterons on \nuc{16}{O},  showing the real (upper panel)
and imaginary (lower panel) DPPs
for specific breakup couplings. The solid lines give the DPP for the L0 case,  the dashed lines are
for the L2 case and the dotted lines are for the L0andL2 case, as defined in the text.  } \end{figure}

In Fig.~\ref{fig3} we compare the DPPs for three cases: L0, L2 and L0andL2, that is 
for breakup to $L=0$ states alone, breakup to $L=2$ states alone and the case when both $ L=0$ 
continuum states and $L=2$ continuum states are coupled to the elastic channel, but (unlike the L02 case) with no
mutual coupling between them. The $L=0$ coupling is overall attractive, while $L=2$ coupling is
overall repulsive, apparently a generic feature, see Ref.~\cite{pang}.  The differences between 
the  L0andL2 and L02 DPPs, that are apparent from a comparison of the solid lines in
 Fig.~\ref{fig2} with the dotted lines in Fig.~\ref{fig3}, reveal the importance of mutual coupling 
between the two continua. The further significance of this is discussed below.

Other generic properties of breakup effects  emerge when we compare 
various characteristics of all the DPPs. The volume integrals of the real and
imaginary parts of the DPPs are calculated by subtracting the volume integrals
$J_{\rm R}$ and $J_{\rm I}$   of the real and imaginary parts of the bare potential, 
from those of the inverted  potentials,  to give $\Delta J_{\rm R}$ and
$\Delta J_{\rm I}$ presented in Table~\ref{dpps}. The volume integrals
are  conventionally defined~\cite{satchler}, with positive
signs indicating attraction or absorption. Table~\ref{dpps} also presents: the
changes in rms radii, $\Delta R_{\rm
R}(\rm rms) $ and $\Delta R_{\rm I}(\rm rms)$, of the real and imaginary potentials; the 
change $\Delta$CS  in reaction cross section due to the coupling; the quantity
$\rho_{\rm I}$ defined below; and the cross section to the breakup states, indicated
as `BU CS' in the tables.
The value of $\Delta J_{\rm I}$ for the L02 case is much less than the sum of the values 
for the L0 and L2 cases, an effect of continuum-continuum coupling. It is generally accepted
that the continuum-continuum coupling is important, but a comparison of the 
L0andL2 and L02 DPPs etc. in Table~\ref{dpps} shows the importance quantitatively.  

\begin{table}[htbp] \caption{For 30 MeV deuterons scattering from \nuc{16}{O},
characteristics of the DPP generated by the L0, L2, L02 and L024 couplings defined in
the text. For case L0andL2 there is coupling to the L0 and L2 
continuum states, but with no coupling between them.
 $\Delta J_{\rm R}$ and $\Delta J_{\rm I}$ are
volume integrals of the real and imaginary DPPs  in  MeV
fm$^3$; $\Delta R_{\rm R}({\rm rms}) $ and $\Delta R_{\rm I}({\rm rms})$
are the changes in rms radius of the real and imaginary terms in fm;  $\Delta$CS 
is the change in reaction cross section due to breakup in mb and  $\rho_{\rm I}$ is
defined in the text. BU CS  is the cross section to the breakup states in mb.} 
\begin{tabular}{lccccccc} \hline 
Coupling &$\Delta J_{\rm R}$ &$\Delta
R_{\rm R}(\rm rms) $& $\Delta J_{\rm I}$ &   $\Delta R_{\rm I}(\rm rms)$ &
$\Delta$CS  & $\rho_{\rm I}$ & BU CS \\ \hline\hline 
L0 & $15.61$ & $-0.0547$ & 26.14 & $0.0196$ & 67.2&2.571 & 63.36 \\ 
L2  & $-24.89$& $-0.0735$ & 35.25& $0.1360$ &79.60&2.258 & 83.30\\ 
L0andL2 & $-14.31$ & $-0.1205$ & 68.34& 0.0765 & 134.10 &1.962& 135.29 \\
L02   & $-9.04$ & $-0.1932$ & 35.42& $0.1190$ & 69.60& 1.965& 83.32 \\ 
L024  & $-11.55$ & $-0.2320$ & 38.11 & $0.1105$& 55.60 &1.459&93.04\\ \hline 
\end{tabular} \label{dpps} \end{table}

\begin{table}[hbt] \caption{For deuterons scattering from \nuc{39}{Ca} or
$^{40}$Ca at indicated laboratory energies in  MeV. All other quantities have the same meaning as elsewhere.}
\begin{center} \begin{tabular}{lcrrrrrrrr} \hline
$E$ (lab.) &Target &Coupling &$\Delta J_{\rm R}$ &$\Delta R_{\rm R}(\rm rms)$& $\Delta J_{\rm I}$& $\Delta R_{\rm I}(\rm rms)$ & $\Delta$CS & $\rho_{\rm I}$& BU CS \\ \hline\hline
26.92& \nuc{39}{Ca}& L02&$ -12.74$&$-0.2050$& 30.23 & 0.1819 &54.7& 1.810 &64.45\\
26.92&\nuc{39}{Ca}& L024 & $-18.10$&$-0.2625$& 28.62& 0.1324 & 26.2 & 0.915& 71.74\\ \hline
30.0&\nuc{40}{Ca}&L02& $-15.35$ & $-0.2396$& 25.69& 0.2217& 62.6& 2.437 &72.18 \\
30.0&\nuc{40}{Ca}&L024 & $-12.32$ & $-0.2411$& 26.32& 0.0835& 34.9& 1.325 &78.35\\ \hline
52 &\nuc{40}{Ca}& L02& $-3.31$ &$-0.1413$ & 25.82 & 0.1688& 97.7 & 2.622 &95.70 \\
52 &\nuc{40}{Ca}& L024& $-3.78$ &$-0.1606$& 27.07& 0.1410 &82.5 & 3.048& 91.64\\ \hline
\end{tabular}
\label{dpps40}\end{center}
\end{table}

\begin{table}[htbp] \caption{For 56 MeV deuterons scattering from \nuc{58}{Ni},
characteristics of the DPP generated by the L0, L2 and L02 couplings defined in
the text.  All other quantities have the same meaning as elsewhere.} 
\begin{tabular}{lccccccc} \hline 
Coupling &$\Delta J_{\rm R}$ &$\Delta R_{\rm R}(\rm rms) $& $\Delta J_{\rm I}$ &   
$\Delta R_{\rm I}(\rm rms)$ & $\Delta$CS & $\rho_{\rm I}$  & BU CS \\ \hline\hline 
L0  & $4.77$ & $-0.042$ & 15.25 & $0.0643$ & 90.4&5.92 & 84.65 \\ 
L2 & $-6.24$& $-0.0628$ & 15.06& $0.2245$ &86.6&5.75 & 88.25\\ 
L02  & $-1.55$ & $-0.141$ & 18.29& $0.1956$ & 86.7&4.74 & 83.62 \\ \hline 
\end{tabular} \label{dppsNi} \end{table}

For the L0, L2 and L0andL2 cases, there is an approximate proportionality between the 
change in reaction cross
section and the change in the volume integral of the imaginary potential. The quantity
$\rho_{\rm I} = \Delta {\rm CS}/ \Delta J_{\rm I}$ presented in Table~\ref{dpps} is 
roughly the same for the L0 and L2 cases, 
but  is distinctly less in the L02andL2, L02 and L024 cases.  

Similar L02 and L024 breakup calculations have 
been carried out for deuterons on \nuc{39}{Ca} and \nuc{40}{Ca} and quantities
characterising the DPPs are given in Table~\ref{dpps40}.
As in the other cases, $L=4$ coupling decreases $\Delta$CS. In 9 of the 11
cases in Table~\ref{dpps} and Table~\ref{dpps40}
the increase in total reaction cross section, $\Delta$CS, is
less than the cross section to the breakup channels, `BU CS' in the table. This indicates that the 
inhibition of other absorptive processes by breakup is quite a general property.
Further indications of the generic nature of breakup effects  can be found in Table~\ref{dppsNi} 
which presents the same characteristics for 56 MeV
deuterons scattering from \nuc{58}{Ni}, see Ref.~\cite{pang}. The magnitudes of
$\Delta J_{\rm R}$ and $\Delta J_{\rm I}$ are less for the higher energy
deuterons, but apart from that, the pattern of changes for the three cases in
common is very similar to what is presented in Table~\ref{dpps}. For 56 MeV deuterons  on \nuc{58}{Ni}, 
the same relationship 
holds between values of $\rho_{\rm I} $ for the L0, L2 and L02 cases 
as shown in Table~\ref{dpps}, although the magnitudes are different. 
Moreover, for both 30 MeV on \nuc{16}{O} and 56 MeV on
\nuc{58}{Ni}, for example, $\Delta J_{\rm R}$ is positive for L0 coupling,
negative but larger in magnitude for  L2 coupling and negative and smallest in
magnitude for  L02 coupling. One difference is that for L02  coupling,
$\Delta$CS  $>$  BU CS   for 52 MeV deuterons on \nuc{40}{Ca} and also for 56 MeV deuterons on
\nuc{58}{Ni}, but the opposite is true for the 30 MeV cases. It might be expected
that the increase in reaction cross section would at least equal the breakup cross
section, but this is only true for the L0 cases and the 52 MeV and  56 MeV case with L02 coupling. 
The effect of breakup in reducing $\Delta$CS  below the breakup cross section is most apparent 
for the L024 cases at  the lower energies. For breakup on \nuc{16}{O} this effect 
is significant:   for  L024 coupling the $L=4$ breakup 
continuum reduces the reaction cross section and at the same time leads to the largest breakup cross section.
Evidently, the coupling to the $L=4$ continuum has reduced the absorption as measured by
$\Delta$CS although the magnitude of $\Delta J_{\rm I}$ shows that the
coupling has increased the effective imaginary potential.

In all cases, both L02 and L024 (but not L0) breakup coupling reduce the real volume integral and increase 
the imaginary volume integral. In all cases breakup coupling reduces the  rms radius of the real potential  
and increases the rms radius of the imaginary potential. This effect on the radial properties of
the potential might be discernible in the systematic  comparison of phenomenological deuteron 
potentials and folding model potentials.

\begin{figure}[htb] \centering
\includegraphics[width=0.6\textwidth,angle=-90]{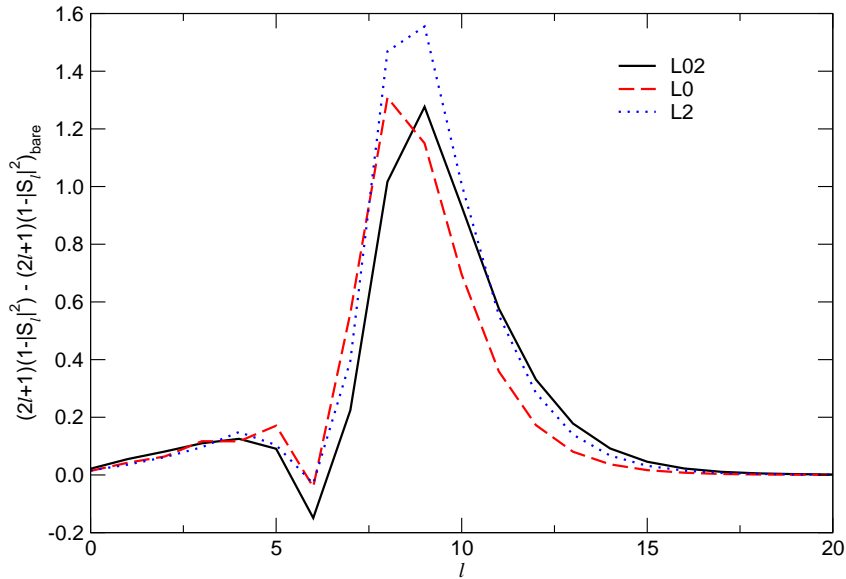}
\caption{\label{fig5}  For 30.0 MeV deuterons on \nuc{16}{O},  the quantity $R(l)$ defined in Eqn.~\ref{R}
is plotted  for the L0 (dashed), L2(dotted) 
and L02 (solid) breakup cases.  } \end{figure}

\begin{figure}[htb] \centering
\includegraphics[width=0.6\textwidth,angle=-90]{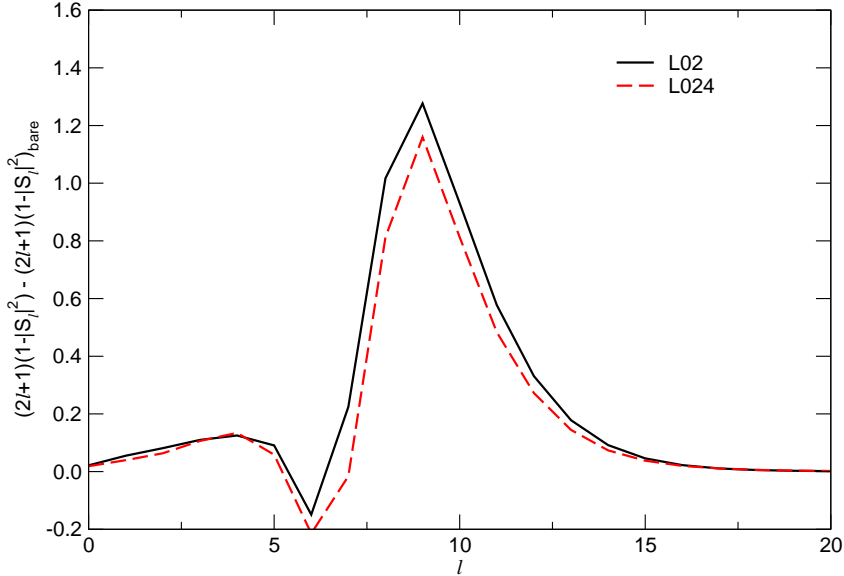}
\caption{\label{fig6}  For 30.0 MeV deuterons on \nuc{16}{O},  the quantity $R(l)$ defined in Eqn.~\ref{R}
is plotted against  $l$ for the L02 (solid line) and  
L024(dashed line) cases.  The partial wave reaction cross section is reduced, for
almost all $l$, by the coupling to the $L=4$ continuum, as also reflected in $\Delta$CS.  } \end{figure}

In order to throw some light on these effects we have plotted the quantity $R(l)$
which can be calculated for the case of any particular breakup coupling:
\beq 
R(l) = (2 l +1)(1- |S_l|^2) - (2l+1) (1- |S_l|^2)_{\rm bare}. \label{R}
\eeq
The subscript `bare' indicates the $S_l$ is for the bare potential. $R(l)$ is a 
measure of the contribution for partial wave $l$ 
to the change in reaction cross section induced by the coupling.
In Fig.~\ref{fig5} a comparison of $R(l)$ for the L0,  L2 and L02 cases shows that the 
extra absorption generated by breakup tends to be at higher $l$ for L2 coupling than
for L0 coupling.  Comparison of $R(l)$ for the L2 and L02 cases shows that the
coupling between the $L=0$ and $L=2$ continua reduces the contribution of 
breakup to the reaction cross section except at the largest $l$ values. This is consistent with the
$\Delta$CS values in Table~\ref{dpps}. For all cases, breakup coupling actually 
decreases the reaction cross section for partial wave $l=6$, a `wrong way' effect as
discussed in Ref.~\cite{rsmprc88}. Fig.~\ref{fig6} shows that inclusion of BU to the $L=4$ 
continuum further reduces the partial wave reaction cross section for almost all partial waves, although 
this coupling does increase the breakup cross section.

\begin{figure}[htb] \centering
\includegraphics[width=0.7\textwidth,angle=-90]{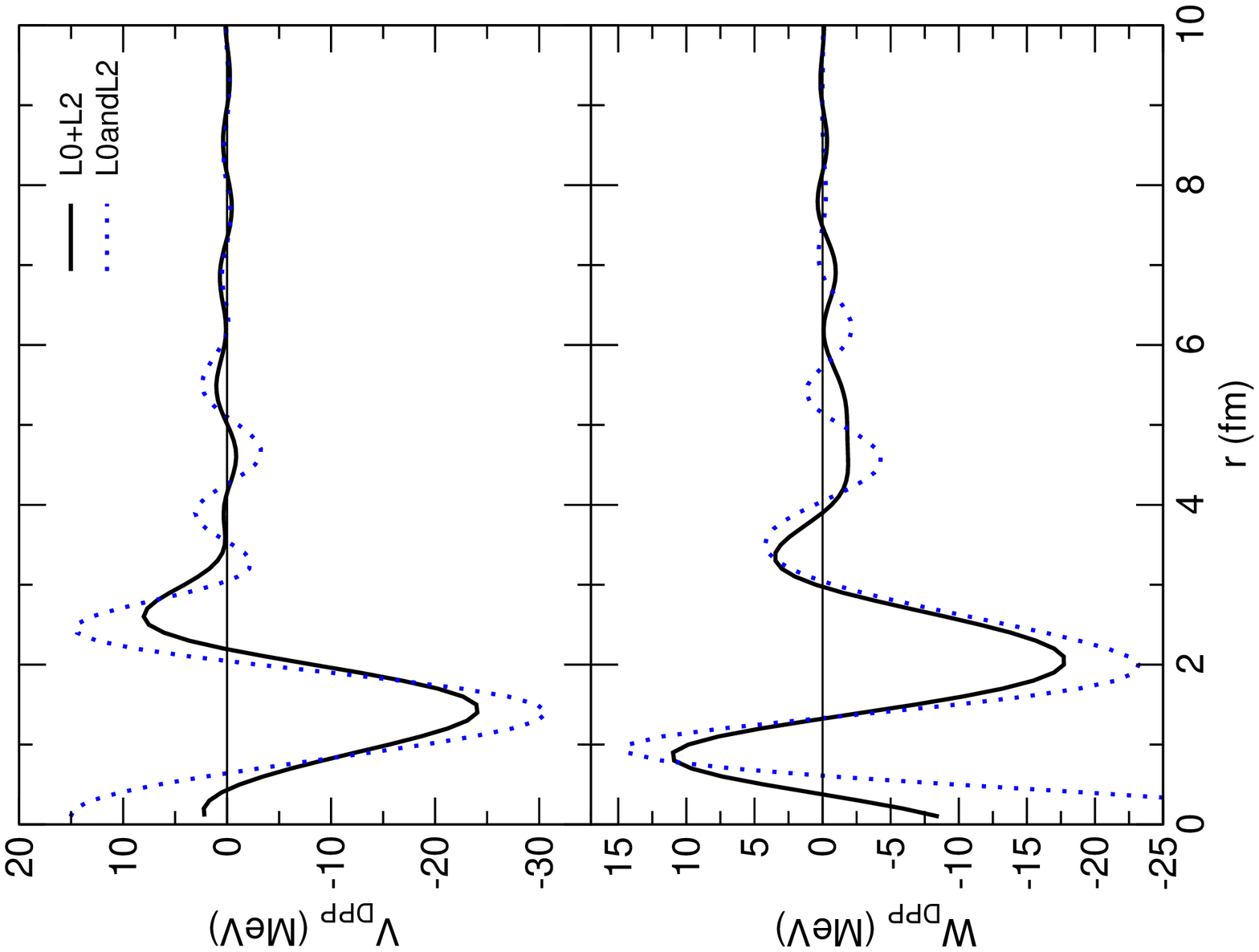}
\caption{\label{fig4}  For 30.0 MeV deuterons on \nuc{16}{O},  showing the real (upper panel)
and imaginary (lower panel) DPPs for specific breakup couplings. The solid lines give the 
sum of the  DPPs for the L0 and L2 cases,  and the dotted lines give the DPP for the 
L0andL2 case.  } \end{figure}

\section{FURTHER IMPLICATIONS OF THE BREAKUP CALCULATIONS}\label{imply}
Apart from the characteristics of the DPPs, there are further implications to be drawn from the
breakup calculations.

\subsection{Evidence for dynamical non-locality}\label{nonloc}
The L0andL2 case is of particular interest: There is no coupling between the L0 channels 
and the L2 channels and, as a consequence, the DPP generated by L0 coupling should
add to the DPP generated by L2 coupling to give the DPP for  L0andL2 coupling.
This additivity rule applies to the underlying non-local DPPs; the formal argument is given in the Appendix.
When, as here, the local equivalent potentials do not add to give the total local 
DPP,  this is evidence for dynamical non-locality of the underlying DPPs. Examples have been presented 
in Refs.~\cite{prc81,prc83} and in the first of these references a simple argument for the non-additivity 
of the local equivalents of non-local potentials was given. 

From Table~\ref{dpps} we see that the sum of the  $\Delta J_{\rm R}$ values for the L0 and L2 cases
is $-9.28$ MeV fm$^3$ whereas it is $-14.31$  MeV fm$^3$ for the L0andL2 case in which there is
no inter-continuum coupling. The corresponding figures for $\Delta J_{\rm I}$ values are 61.39 MeV fm$^3$
and 68.34 MeV fm$^3$, respectively.
This is evident also from the point-by-point DPPs
presented in Fig.~\ref{fig4} which compares the (local equivalent) DPP for the 
L0andL2 case with the sum of the local DPPs for the L0 and L2 cases. There is
a substantial difference.

It will be noticed from Table~\ref{dpps} that the breakup cross section for the L0andL2 case, 135.29 mb, 
is not equal to the sum of the L0 and L2  breakup cross sections,  63.36 + 83.30 = 146.66 mb. 
In fact, in the L0andL2 case, the individual cross sections for the L0 and L2 channels are 56.16 mb 
and 79.13 mb, respectively. The L0 cross section, but not the L2 cross section, is modified by the coupling of 
the other channel to the elastic channel, in the absence of direct coupling.
Because the elastic channel wave function is influenced by the sum of the DPPs, this does not conflict 
with the fact that the non-local DPPs add.  This  indirect influence is an argument for comparing results 
using non-local and local equivalent distorting potentials in direct reactions. The Appendix gives a formal 
account of these effects.

\subsection{Consequences of interactions between breakup continua}\label{interactions}
It is conspicuous that when the monopole and quadrupole excitation channels are inter-coupled,
as in the L02 case, neither the total reaction cross section nor the breakup cross section increases 
much as a result.
Indeed, $\Delta$CS for the L02 case is less than for the L2 case. Can we explain 
why the additional inclusion of $L=0$ breakup reduces the reaction cross section? A clue is 
one aspect of non-locality as follows:
The behavior of $|S_l|$ for deuterons is intermediate between that for nucleons, for which $|S_l|$ 
does not become very small for the lowest $l$,  and the behavior for heavier particles.
For the lowest values of $l$,  $|S_l|$ becomes progressively smaller as the projectile mass increases 
see Ref.~\cite{cook,grs2} (\nuc{6}{Li} and \nuc{7}{Li} being exceptions). For deuterons, the
relatively large $|S_l|$ for low $l$ is consistent with the results of the notch test 
and suggests that there is a substantial probability of a deuteron returning to its ground state 
after its encounter with the nucleus.  
The excitation out of its ground state, out of the elastic channel, and then back, is a non-local
effect, see Austern~\cite{austern}. The excitation of the deuteron from its ground state can
be considered a temporary distortion and when monopole and quadrupole continua 
are coupled together, the total excitation of the pair is roughly the excitation of each alone.
The two nucleons evidently penetrate the target 
nucleus more like free  nucleons than like nucleons in a more tightly bound heavier projectile nucleus. 
It appears that the fragility of the deuteron actually facilitates its survival as it interacts with the nucleus,
as suggested by Rawitscher long ago~\cite{rawit67}.

\subsection{Further generic properties}\label{generic} 
In Section~\ref{dpdt} regularities in the DPPs due to deuteron breakup were identified as generic, but
they seem to apply to more general cases of projectile breakup.
Certain characteristics  of DPPs due to breakup have previously been identified as universal~\cite{arPLB169}, 
occurring with the breakup of both~\nuc{2}{H} and \nuc{6}{Li} projectiles.  For deuterons, they apply
for much heavier target nuclei than~\nuc{16}{O}. In all cases the DPPs had strong undulations 
and the undulations in the imaginary DPP generally
involve radial ranges where the DPP is emissive or nearly so.  Also, in all cases L0 breakup generated weak surface attraction, and also attraction within the nucleus, whereas L02 breakup led to much stronger surface repulsion, with attraction within the nucleus, although the radial form is different for L0 breakup.  These features appear in varying degrees in the DPPs found here.  As further discussed in Section~\ref{reTELP},  such undulations can be 
associated with $l$ dependence, particularly when the coupling processes lead to substantially different
effects for $l$ less than or greater than the value for which $|S_l| \sim \hlf$. The particularly strong contribution 
of low-$l$ partial waves to the scattering of deuterons from \nuc{16}{O}, as revealed by a notch test,  makes deuteron scattering from \nuc{16}{O} at 30 MeV susceptible to this effect.

\section{TELP inversion: evaluation and implications}\label{reTELP}
\subsection{Evaluation of TELP inversion}
The use of  a weighted TELP inverted potential  as the SRP for inversion, when using the Imago inversion code, 
has a useful by-product. That by-product is all the information required to  determine the volume integrals and other characteristics of the DPP  calculated from  the TELP potential itself. These characteristics are presented in Table~\ref{ttelp} and can be compared with the same characteristics presented in Table~\ref{dpps} for DPPs calculated using exact S-matrix inversion. The differences are large,  particularly for the L02 and L024 cases. In both cases $\Delta J_{\rm I}$ is higher while at the same time $\Delta$CS is much lower.  For all cases$|S_l|$ calculated from the  weighted TELP is greater than that produced by the coupled channel code for all $l$ greater than that from which
$|S_l| \sim \hlf$. This tends to reduce $\Delta$CS.  As a result,  $\rho_{\rm I}$ is, for the L02 and L024  cases respectively, about a half and a third of the values for the S-matrix inverted potentials. This is consistent with the fact that
$\Delta R_{\rm I}(\rm rms)$ is much too low for TELP potentials, suggesting that the imaginary potential is shifted
inwards compared to the exact S-matrix inverted potential.

\begin{table}[htbp] \caption{For 30 MeV deuterons scattering from \nuc{16}{O},
characteristics of the DPP, generated by the L0, L2, L0andL2, L02 and L024 couplings defined in
the text.  These values relate to the potentials calculated by TELP inversion, and all 
other aspects of the Table are as for Table~\ref{dpps}. } 
\begin{tabular}{lccccccc} \hline 
Coupling &$\Delta J_{\rm R}$ &$\Delta
R_{\rm R}(\rm rms) $& $\Delta J_{\rm I}$ &   $\Delta R_{\rm I}(\rm rms)$ &
$\Delta$CS  & $\rho_{\rm I}$ & BU CS \\ \hline\hline 
L0 & $6.05$ & $-0.0684$ & 19.74 & $-0.0230$ & 43.80&2.219 & 63.36 \\ 
L2  & $-9.99$& $-0.0903$ & 42.40& $0.0916$ &55.50&1.309 & 83.30\\ 
L0andL2 & $-3.63$ & $-0.1618$ & 61.65 & 0.0589 & 87.3&1.416& 135.29 \\
L02   & $-1.74$ & $-0.1772$ & 42.92& $0.0375$ & 39.20& 0.913& 83.32 \\ 
L024  & $-4.47$ & $-0.2016$ & 42.11 & $0.0024$& 23.30 &0.553&93.04\\ \hline 
\end{tabular} \label{ttelp} \end{table}

Some  qualitative properties remain  the same, e.g. $\Delta J_{\rm R}$ is positive for the L0 case and
negative for the L2 case, but the magnitudes are very different. Therefore,  conclusions 
from quantities such as those presented in Table~\ref{dpps} could not be reliably obtained from TELP
potentials. The much lower values of $\Delta$CS presented in Table~\ref{ttelp} would greatly
exaggerate the relationship noted above between $\Delta$CS  and the cross section to breakup channels.

The elastic scattering angular distributions  for the TELP inverted potentials differ considerably from the corresponding coupled channel elastic scattering angular distributionss,  particularly at backward angles. For the L02 case at 150$^{\rm o}$ the angular distribution corresponding to the TELP potential is a factor of 2 lower than the true value.  The S-matrix inverted  potentials always precisely reproduce the coupled channel elastic scattering angular distributions.

The TELP potentials reflect the non-additivity of the local equivalent DPPs in the L0andL2 case where the sums
of  $\Delta J_{\rm R}$, $\Delta J_{\rm I}$  and $\Delta$CS  in the L0 and L2 cases in Table~\ref{ttelp}  are respectively $-3.94$ MeV fm$^3$, 62.14 MeV fm$^3$ and 99.30 mb  compared with the values in the L0andL2 line.

\begin{figure}[htb] \centering
\includegraphics[width=0.7\textwidth,angle=-90]{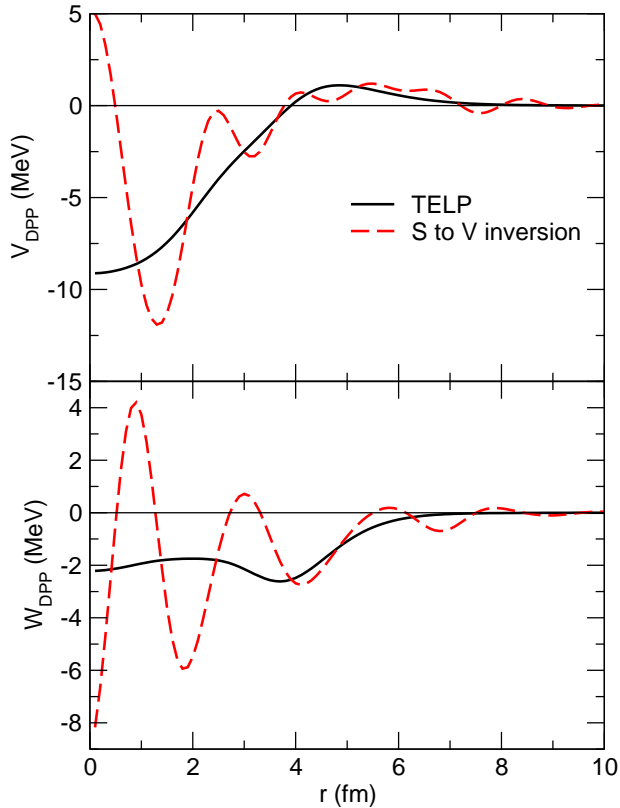}
\caption{\label{telp-comp}  For 30.0 MeV deuterons on \nuc{16}{O},  comparing the exact and TELP inverted potentials for the case with L024 coupling. The solid lines give the TELP potential and the dashed lines the exact S-matrix inverted potential. The real potential is in the upper panel and the imaginary part in the lower panel. } \end{figure}

\begin{figure}[htb] \centering
\includegraphics[width=0.7\textwidth,angle=-90]{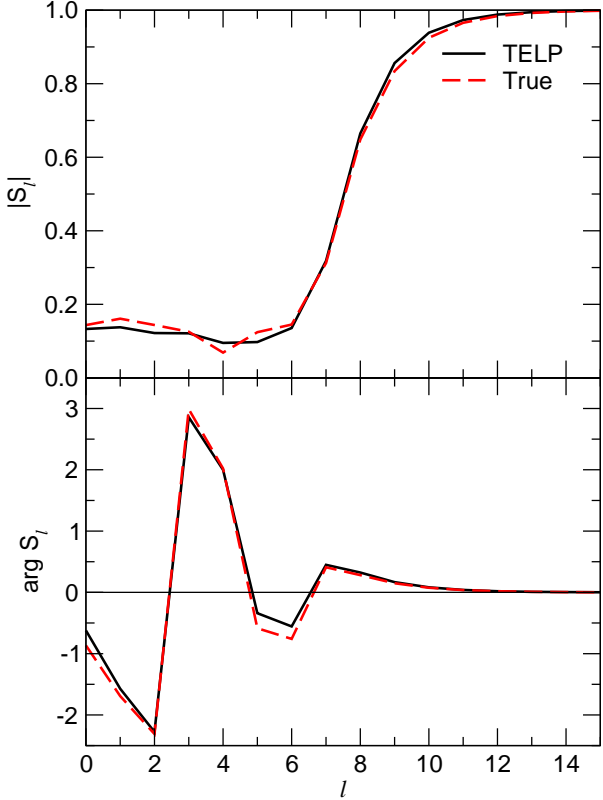}
\caption{\label{telp-smes}  For 30.0 MeV deuterons on \nuc{16}{O},  comparing the CDCC $S_l$  with $S_l$ calculated from the TELP, both for L024 coupling. The solid lines give the TELP values and the dashed lines the CDCC values, with $|S_l|$ in the upper panel and $\arg S_l$ in the lower panel.} \end{figure}

\subsection{Conclusions drawn from TELP potentials}\label{telp-conc}
The TELP inverted potentials are smooth and do not have the undularity, 
evident in Fig.~\ref{fig1},  of the S-matrix inverted potentials. Figure~\ref{telp-comp} presents a direct comparison of the TELP and exact DPPs for the L024 case. Figure~\ref{telp-smes} compares $S_l$ calculated 
from the TELP potential with that directly from the CDCC 
calculation. It reveals an alternation of the CDCC $S_l$ about the TELP $S_l$:  for example 
$ |S_l|_{\rm CDCC}  - |S_l|_{\rm TELP}$ is $\le 0$ for $ l \ge 6$ and for $l=4$ but $|S_l|_{\rm CDCC}  - |S_l|_{\rm  TELP}> 0$ for all other $l$. The deviations of the L0, L2 and L02 TELP potentials follow very similar patterns to that for the L024 TELP potential, shown in Fig.~\ref{telp-smes} (the L0 case has larger differences for low values of $l$). 

The undularity of the S-matrix equivalent inverted potential appears to reflect the $l$-dependent difference of Fig.~\ref{telp-smes}.  Apart from $l=4$, the difference in $|S_l |$ is between high and low values of $l$; the difference in $\arg S_l$ varies in a similar way. For both L02 and L024 cases, the TELP $|S_l|$ is too large for high $l$ and too low for low $l$. Recalling the $2 l + 1$ weighting factor,
this accounts for the fact that the $\Delta$CS values in Table~\ref{ttelp} are considerably less than the correct values
in Table~\ref{dpps}, affecting the relationship with breakup cross section that was discussed above.  In Ref.~\cite{ldep} it is shown how a small $l$-dependent factor applied to an $l$-independent S-matrix generates undulations in the corresponding inverted  potential. Finally we remark that the undularity in the S-matrix inverted potentials, as seen in Figure~\ref{telp-comp}, is not sporadic, but follows a consistent pattern in all examples of breakup effects that we have studied.

Although TELP inversion is imperfect, \emph{qua} inversion, it has thrown light on the the $l$ dependence of the deuteron OMP, and also provides a useful SRP for inversion in difficult cases.

\section{DISCUSSION AND CONCLUSIONS}\label{conclusions} 
The excitation of a deuteron as it interacts with a target nucleus generates a contribution 
to the deuteron-nucleus potential that is both non-local and $l$ dependent. The conventional 
OMP that is employed to describe  the elastic scattering of deuterons from nuclei is,
however, both local and $l$ independent. Such phenomenological  local potentials are widely regarded as
giving `satisfactory' fits to elastic scattering data, although the requirement for precise fits
results in potentials with unusual features, see for example Ref.~\cite{ermer}.  Local potentials,
often with globally fitted parameters, are widely used in the analysis of direct reactions  
involving deuterons leading to spectroscopic information. The primary goal of the present work,
of which this paper presents the first part,
is to gain an understanding of the consequences, for direct reactions, of the fact that
a conventional  OMP is the local equivalent of a potential that is actually nonlocal.
We emphasise that the subject
is dynamical non-locality, associated with the excitation of one of the interacting nuclei
(the projectile, in this work), and not the non-locality arising from knock-on exchange processes
which, for the particular case of ($d,p$) reactions, can be allowed for, see Ref.~\cite{nunes}.

In order to determine the effects of dynamical non-locality for general direct reactions involving
deuterons, we require the exact local equivalents to the non-local potentials generated
by the breakup of the deuteron as it interacts with the target nucleus.
The initial motivation for the work reported in this paper was the determination of such local 
potentials. However, a number of findings have emerged that are of independent interest.

The interaction and breakup of `fragile' nuclei is a subject with various aspects;  for example,
the effect of breakup on fusion processes   has been exhaustively studied, for a recent
review see Ref.~\cite{canto}. The contribution of breakup to the departure from folding model
systematics was an early motivation for CDCC studies which were mostly devoted to the effect
on the interaction in the surface region. Nevertheless the study of the DPP arising from breakup coupling 
reveals a number of aspects of general significance. In particular, we note that in the case we study,
notch tests reveal that the interest in the DPP is not confined to the surface region.
For deuterons interacting with \nuc{16}{O} the
coupling modifies the potential over almost the whole radial range, and in ways which are of intrinsic
interest for the understanding of inter-nuclear interactions. The dynamically generated interaction 
is not smooth, and a comparison with earlier work suggests that there exist generic properties that apply 
widely to DPPs arising from coupling to breakup channels.  Studying breakup effects in the context of deuteron scattering has  the advantage that, considered as a cluster nucleus, its structure is unique, something that is 
not true of  many other nuclei for which the effects of breakup processes have been studied, 
e.g.\ Ref.~\cite{kmb}. 

Among the facts that have emerged, some are not particularly surprising. An example is the fact that
coupling to $L=2$ breakup states contributes to $|S_l|$ for higher values of $l$ than breakup to the $L=0$ 
continuum. What might be unexpected is the fact that the inclusion of the $L=4$ continuum 
\emph{reduces} the absorption from the elastic channel, or the fact that the inclusion of $L=0$ 
and $L=2$ breakup together, with full mutual coupling, can make a smaller contribution to the total reaction
 cross section than $L=2$ coupling alone.

The local equivalent potentials representing the DPPs have a pattern of undulations which indicates
that the underlying DPP is $l$ dependent as well as non-local.  The undulations
have general features similar to those found in breakup calculations for 56 MeV deuterons on \nuc{58}{Ni},
as well as for deuterons on \nuc{39}{Ca} and \nuc{40}{Ca} over a range of energies. It is also
found that breakup coupling systematically reduces the rms radius of the real part of the potential and increases
the rms radius of the imaginary part, compared to the bare folding model potential.
These apparently generic properties  could be studied with precision elastic scattering experiments
together with precision model-independent fitting. How much these effects would survive the inclusion of
other reaction processes, such as coupling to mass-1 and mass-3 channels, remains for future
studies, as does the full representation of spin effects. The undulatory potentials can not represent an 
$l$-independent potential since such a potential $V(r)$ must have a zero derivative at $r=0$  since otherwise 
the potential as a function in 3 dimensions would have a cusp at the nuclear center.

The occurrence of undulations deserves experimental study. In cases such as that reported  in Ref.~\cite{ermer},
where precise, wide angular range data are fitted exactly using model independent methods, such features
do appear. It is likely that if more precise wide range data were fitted to the same standard that is normal for electron scattering analyses, then  such features would commonly be found. This could then provide an indirect method of
exploring the possible $l$-dependence of the nuclear OMP.   The systematic  undulatory properties of the local potentials are consistent with previous CC plus inversion results, and are more indirectly supported by the results of model independent fitting~\cite{ermer}.  Undularity cannot be ignored and should be considered seriously as a property of the deuteron-nucleus potential.

The  CDCC calculations presented an opportunity to evaluate the weighted TELP 
procedure for inversion. We find that for the present case it would not provide
a reliable method for studying the properties of the DPP.  The $l$-dependent differences between 
$S_l$ direct from the CDCC calculation and $S_l$ from the TELP  potential 
suggest why the inverted potential  is undulatory, and also why the TELP reaction cross section is incorrect.
This can be  seen from the $\Delta$ CS values  in Table~\ref{ttelp} which are too low. 

The local potentials derived here are applied in a subsequent paper, Ref.~\cite{paper2}, to the study
of the effect of dynamical non-locality on transfer reactions. However, the present work has already shown,
through the non-additivity of local equivalent DPPs, that breakup coupling leads to appreciable non-local effects.

\section{Appendix: Adding non-local DPPs}
We present a simplified model demonstration that formal non-local DPPs arising
from coupling to channels that are coupled to the elastic channel but not to each other
add  to give the total non-local DPP.  We consider a spinless projectile on a spinless target so the total 
conserved angular momentum is 
the orbital angular momentum. The orbital angular momentum operator is implicit in the 
kinetic energy operator $T$  in the coupled channel equations, with channel 0 the elastic channel.
We first consider the example of two spinless states:

\beq (T + V_{00}(r) -E_0) \psi_0(r) =  - V_{01}(r) \psi_1(r) - V_{02}(r) \psi_2(r),  \label{E1}\eeq

\beq (T + V_{11}(r) -E_1) \psi_1(r) =  - V_{10}(r) \psi_0(r) - V_{12}(r) \psi_2(r),  \label{E2}\eeq

\beq (T + V_{22}(r) -E_2) \psi_2(r) =  - V_{20}(r) \psi_0(r) - V_{21}(r) \psi_1(r).  \label{E3}\eeq

If there is no coupling between channels 1 and 2, Eqs~\ref{E2} and~\ref{E3} become
 Eqs~\ref{E2a} and~\ref{E3a}:
\beq (T + V_{11}(r) -E_1) \psi_1(r) =  - V_{10}(r) \psi_0(r),   \label{E2a}\eeq

\beq (T + V_{22}(r) -E_2) \psi_2(r) =  - V_{20}(r) \psi_0(r).  \label{E3a}\eeq

We can rewrite the last two equations, defining $G_1$ and $G_2$ as:
\beq \psi_1 = \frac{1}{E^+_1 -T -V_{11}}V_{10}\psi_0 \equiv G_1 V_{10} \psi_0 \label{E2b} \eeq
and 
\beq \psi_2 = \frac{1}{E^+_2 -T -V_{22}}V_{20}\psi_0 \equiv G_2 V_{20} \psi_0. \label{E2bb} \eeq

The equation for the elastic channel wave function $\psi_0$ is therefore
\beq (T + V_{00}(r) -E_0) \psi_0(r) =  - V_{01}G_1 V_{10} \psi_0 - V_{02}G_2 V_{20} \psi_0  \label{ECC}\eeq

so the effective elastic channel potential is
\beq V_{00} + V_{01} G_1 V_{10}  + V_{02} G_2 V_{20} \equiv V_{00} + {\rm DPP}_1 + {\rm DPP}_2. \label{Esum} \eeq
which is to say that the (non-local and $l$-dependent) DPPs due to the coupling to channel 1 and to channel 2 add to 
give the total (non-local and $l$-dependent) DPP. Note, however, that $\psi_1$ is affected by the coupling in channel 2 through the effect on the elastic channel $\psi_0$ so that the inelastic cross sections in each of channel 1 and channel 2 may be strongly dependent on the coupling in the other channel.

The important point here is that although the total DPP is the sum of the DPPs separately due to the excitations in channels 1 and 2,
the local and $l$-independent representation of this DPP, i.e. the local potential that gives in a single channel
calculation the same $\psi_0$ in the asymptotic region, and hence the same elastic S-matrix $S_l$, as the coupled equations,
will certainly not be a sum of the local representations of  $ {\rm DPP}_1$ and $ {\rm DPP}_2$. The local equivalent of the sum
of two non-local potentials is not the sum of the local equivalents of each potential, see Ref.~\cite{prc81}. 

It is straightforward to see that the above all holds when equations ~\ref{E2a} and~\ref{E3a} become sets of coupled equations
with no coupling between each set:
\beq(T +V_{ii} -E_i) \psi_i = - \sum_j V_{ij}\psi_j -V_{i0}\psi_0 \label{E2c} \eeq
\beq(T +V_{mm} -E_m) \psi_m = - \sum_n V_{mn}\psi_n -V_{m0}\psi_0 \label{E2d} \eeq
in which case Eqn.~\ref{E1} becomes
\beq (T + V_{00}(r) -E_0) \psi_0(r) =  - \sum_iV_{0i}(r) \psi_i(r) - \sum_mV_{0m}(r) \psi_m(r).  \label{E1c}\eeq

The formal (vector) solution to Eqn.~\ref{E2c} can be written in terms of the coupled Green function for
\beq (T +V_{ii} -E_i) \psi_i + \sum_j V_{ij}\psi_j =0 \eeq
i.e.
\beq G_{ij} \equiv \frac{1}{E^+ - H_{ij}}\eeq so again we get the total potential:
\beq V_{00} + \sum_{ij} V_{0i} G_{ij} V_{j0}  + \sum_{mn} V_{0m}G_{mn} V_{n0}. \eeq
Thus, the non-local DPP arising from the coupling of the elastic channel to a set of channels $i, j, \ldots$, which are coupled together,
adds to the non-local DPP arising from coupling of the elastic channel to a set of channels $m, n, \ldots$ to give the total non-local
DPP providing there is no coupling between the channels $i, j, \ldots$ and the channels 
$m, n, \ldots$ . But, there is no reason to suppose that the local equivalents add in the same way to give the total local equivalent DPP.

The local DPPs are far from being additive in the present deuteron breakup cases, and this can be taken as an indication of the dynamical non-locality of the separate DPPs. Again, although the coupling in one set of channels has no influence on the (non-local, $l$-dependent) DPP arising from the other set
of channels, coupling in each set \emph{does} affect inelastic cross sections in the alternate set of channels. 

In some circumstances the local equivalent DPPs do add quite closely to give the total local DPP, and the comparison might be informative.  A case where the local equivalent  DPPs do add very closely is that of  80 MeV \nuc{16}{O} scattering from \nuc{208}{Pb}~\cite{PL161}. There were two classes of local DPPs: (i)  due to inelastic excitation, and, (ii)  due to particle transfer. There was no mutual coupling between the collective and transfer processes. That case differs from the present deuteron case in a number of respects: the projectile wavelength is much shorter, and the  coupling, and hence the DPPs, are confined to the surface region.  In addition, the bare potential was almost purely real in the active radial range which is probably why the real DPPs were predominantly attractive in the surface region, unlike the deuteron breakup DPPs.

\section{Acknowledgment} The authors are  very grateful to Ian Thompson  for modifications of  his powerful FRESCO code enabling the L0andL2 calculations.

\end{document}